\documentclass[pmlr]{jmlr}

\usepackage{longtable}
\PassOptionsToPackage{table,xcdraw}{xcolor}
 \usepackage{siunitx}

\theorembodyfont{\upshape}
\theoremheaderfont{\scshape}
\theorempostheader{:}
\theoremsep{\newline}

\jmlrvolume{1}
\jmlryear{2024}
\jmlrworkshop{AAAI 2025 Workshop iRAISE}

\title[Few-Shot GPT Models vs. BERT for Open-Response Equity Assessment]{Comparing Few-Shot Prompting of GPT-4 LLMs with BERT Classifiers for Open-Response Assessment in Tutor Equity Training}

 \author{\Name{Sanjit Kakarla} \Email{sanjit.kakarla@gmail.com\\
 \Name{Conrad Borchers} \Email{cborcher@cs.cmu.edu}\\
 \Name{Danielle Thomas} \Email{drthomas@cmu.edu}\\
 \Name{Shambhavi Bhushan} \Email{shambhab@andrew.cmu.edu}\\
   \Name{Kenneth R. Koedinger} \Email{koedinger@cmu.edu}\\
   \addr Human-Computer Interaction Institute \\Carnegie Mellon University\\5000 Forbes Ave.\\Pittsburgh, PA 15213, USA}}

\begin{document}

\maketitle

\begin{abstract}
Assessing learners in ill-defined domains, such as scenario-based human tutoring training, is an area of limited research. Equity training requires a nuanced understanding of context, but do contemporary large language models (LLMs) have a knowledge base that can navigate these nuances? Legacy transformer models like BERT, in contrast, have less real-world knowledge but can be more easily fine-tuned than commercial LLMs. Here, we study whether fine-tuning BERT on human annotations outperforms state-of-the-art LLMs (GPT-4o and GPT-4-Turbo) with few-shot prompting and instruction. We evaluate performance on four prediction tasks involving generating and explaining open-ended responses in advocacy-focused training lessons in a higher education student population learning to become middle school tutors. Leveraging a dataset of 243 human-annotated open responses from tutor training lessons, we find that BERT demonstrates superior performance using an offline fine-tuning approach, which is more resource-efficient than commercial GPT models. We conclude that contemporary GPT models may not adequately capture nuanced response patterns, especially in complex tasks requiring explanation. This work advances the understanding of AI-driven learner evaluation under the lens of fine-tuning versus few-shot prompting on the nuanced task of equity training, contributing to more effective training solutions and assisting practitioners in choosing adequate assessment methods. 
\end{abstract}

\section{Introduction}

Assessing learners' responses to open-ended questions in ill-defined domains is an area of ongoing research and a novel educational task for language models. In education, ill-defined domains include situational judgment testing, involving assessing complex scenarios that lack clear right or wrong answers. These domains are constructs where the ``correct'' answer to a question is not well-defined and often require nuanced reasoning and decision-making, such as identifying the most appropriate response to students experiencing inequities or motivational challenges \citep{thomas2023tutor}. Traditional assessment methods often involve humans scoring responses based on complex rubrics given the context-dependent nature of such tasks, creating a need for novel scalable, automated approaches to evaluation \citep{vercellotti2021beyond}.

Advancements in natural language processing (NLP) offer promising solutions for automating assessments in these contexts. Earlier-generation models, such as Bidirectional Encoder Representations from Transformers (BERT) \citep{devlin2018bert}, RoBERTa \citep{liao2021improved}, and T5 \citep{carmo2020ptt5}, provide efficient, low-cost, and computationally resource-light options, while large language models (LLMs) like GPT-4 models and Gemini excel in capturing contextual nuances but are comparatively resource-intensive \citep{brown2020language}. As state-of-the-art LLMs, OpenAI GPT models have access to a wide range of real-world knowledge represented in their training corpus, which could show satisfactory accuracy with human-like reasoning abilities when prompted with few examples \citep{brown2020language}. 

Despite the promise of LLMs, their performance and cost-effectiveness in real-world educational contexts remain underexplored, particularly in tasks requiring subtle interpretation of human responses. Legacy models like BERT, on the other hand, are more limited in real-world knowledge but can be fine-tuned offline based on a small set of human-annotated examples, being comparatively computationally lightweight and open-source, meaning that BERT models can be downloaded, inspected, and used by researchers and practitioners with stable versioning. This study addresses these gaps by comparing the performance of BERT, GPT-4o, and GPT-4-turbo in assessing responses to situational challenges in advocacy-focused training lessons. GPT-4.o was used as it is substantially more cost-effective compared to GPT-4, although we aimed to examine both these models on their human-like reasoning abilities within the ill-defined domain of equity training. Using a dataset of 291 lesson completions from 243 college students learning to become middle school tutors, we evaluate these models on their ability to assess their performance by capturing the nuances of open-ended responses. This research contributes to understanding how AI tools can enhance evaluation in ill-defined domains, offering practical insights into selecting appropriate models for diverse educational applications. To this end, we investigate the following research question:

\textbf{RQ1:} What is the comparative performance between legacy natural language processing models such as BERT and contemporary models like GPT-4o and GPT-turbo in assessing learners' open responses within scenario-based equity training? 
\vspace{-0.4cm}
\section{Related Work}
\subsection{Smaller-scale Early-generation Language Models in Education} Early-generation language models have been extensively used in evaluating learner performance through open response grading tasks, evaluating responses that do not have predefined answers. These models have focused on prioritizing efficiency and task-specific optimization after learning from vast pools of training data from the open web \citep{kaddoura2022systematic}. BERT has played a key role in applying NLP to assessment. BERT has shown its capacity to analyze open-ended responses and other unstructured forms of data due to its bidirectional text understanding, gaining an understanding of all the words in a text body, rather than only analyzing text backward as N-Gram models do \citep{devlin2018bert}. BERT models have excelled in assessment tasks such as automated grading in science education, where they have been fine-tuned to evaluate student understanding of scientific phenomena and the application of experimental principles \citep{zhai2024ai}. Variant models including RoBERTa, XLnet, and T5 have offered incrementally improved performance \citep{liao2021improved, carmo2020ptt5}. Employing these machine learning-based models in the process of assessing human responses saves time and reduces costs \citep{zhai2020applying}. Due to increasing portability to be deployed on different computational resources, these models have become more suitable for local deployments, making them ideal for use in environments with limited educational infrastructure. For instance, T5 has allowed NLP tasks to evolve into a text-to-text framework, simplifying its use in assessing tutor responses \citep{bird2023chatbot}.

\vspace{-0.3cm}
\subsection{LLMs in Education} Large-language models include more parameters than traditional transformer-based models like BERT, allowing for substantially increased model training and a knowledge base from its large training corpus to grade responses. LLMs such as GPT-4 have strong reasoning abilities, which hold potential for use in contextual judgment tasks previously only accessible to humans \citep{kalyan2023survey}. Emerging evidence points to these LLM reasoning abilities allowing for handling nuanced educational tasks such as real-time feedback generation and timely tutor evaluation of specific competencies such as reacting to student errors \citep{lin2024can, kakarla2024using}. Prompt engineering strategies — the ability to craft prompts for generating adequate responses for tasks — can enhance additional flexibility within LLM responses. These improvements, however, come at the costs of computationally heavy training, API use costs, changing model versions over time, and external model deployment issues, such as temporary downtime. Therefore, we are comparing the lightweight, legacy BERT model with GPT models by OpenAI.

\vspace{-0.4cm}
\section{Method}

Our dataset comprised open-response text responses of undergraduate tutors in the Northeastern United States, employed as paid tutors for a remote tutoring organization aiding middle-school students with corresponding human-labeled scores from two lessons: \textit{Avoiding Unconscious Assumptions} (N = 79 human-labeled responses) and \textit{Helping Students Manage Inequity} (N = 164 human-labeled responses). These lessons and their questions were created by a university research team aligned with research-supported, real-world competencies of effective tutor equity training \citep{thomas2024tutors}. The lesson was delivered to tutors via an online learning platform. Label imbalance in these annotated samples ranged from 41\% to 43\% across prediction tasks, which is moderate, with the exception of generating dialog in the Helping Students Manage Inequity lesson, standing at 85\%, which we address through stratification of cross-validation folds. Both sample sizes were determined based on convenience and data availability at the time of writing this manuscript. For each lesson, tutors are tasked to predict and explain the appropriate research-supported approach. Explained responses require learners to explain why they would say something to a specific student, while predicted responses require learners to generate open-ended text to represent real dialogue they would have with students during tutoring. The predict question for a sample scenario within the \textit{Helping Students Manage Inequity} lesson is shown in Figure 1. The Explain question, shown in Figure 2,  tasks tutors to explain the rationale behind their response.

Each response was graded by human coders in a binary fashion, meaning that the response was either correct and adequate or incorrect and inadequate. A sample response for the situation depicted in Figure 1 on responding to student inequity is \textit{``I would provide him with other possible
ways to access the internet that helps him
do the homework equally. For example, he
could do the homework at school.''} Human coders graded this response as incorrect, given that it is not mindful of inequitable access differences to resources (see digital appendix linked below in this section for full rubrics). Two human coders achieved an inter-rater reliability of Cohen's $\kappa$ of 0.88 and 0.87 for the \textit{Avoiding Unconscious Assumptions} lesson (based on 79 tutor responses) and 0.86 and 0.93 for the \textit{Helping Students Manage Inequity} lesson (based on 164 tutor responses), which indicates excellent reliability \citep{warrens2015five}.

\begin{figure}[h!]
        \centering
        \includegraphics[width=0.65\textwidth]{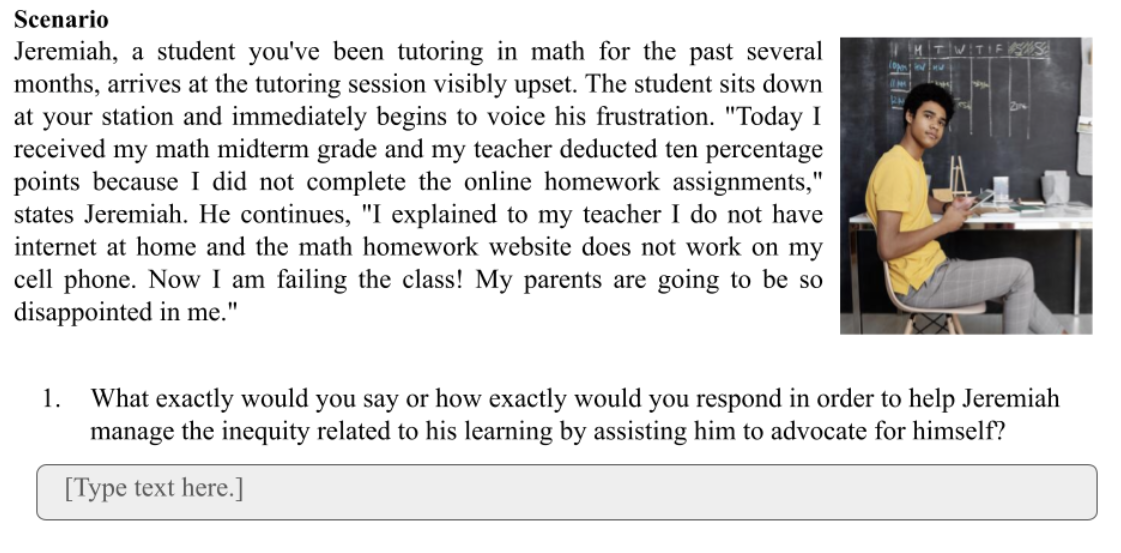}
        \caption{Predict Scenario for the \textit{Helping Students Manage Inequity} Lesson.} 
        \label{fig:1}
    \end{figure}

\begin{figure}[h!]
        \centering
        \includegraphics[width=0.65\textwidth]{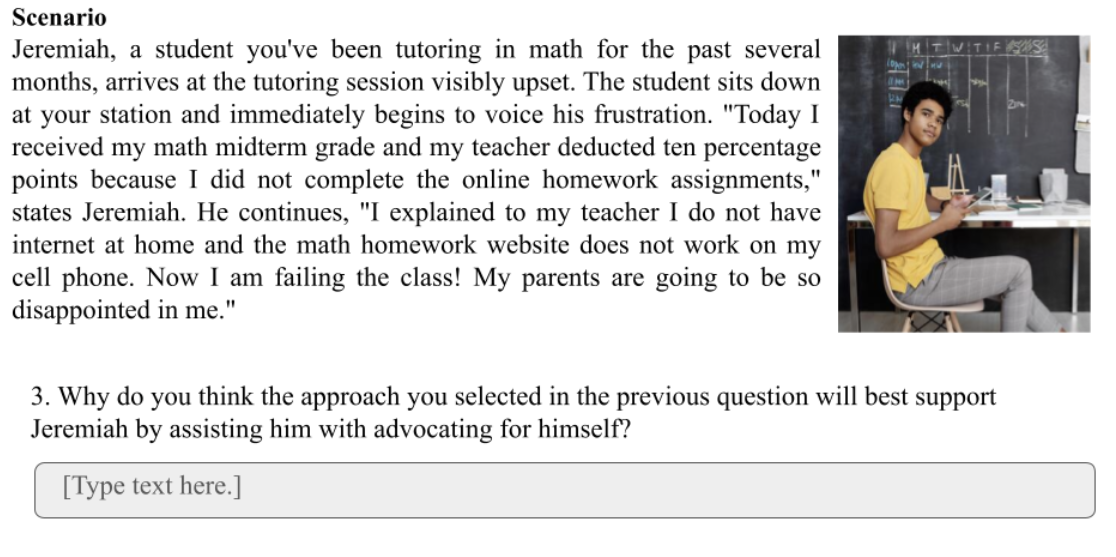}
        \caption{Explain Scenario for the \textit{Helping Students Manage Inequity} Lesson} 
        \label{fig:2}
\end{figure}

For evaluation, human-labeled data were split into training and validation subsets according to stratified 5-fold cross-validation. Given the small overall sample size, no hold-out test set evaluation was performed. Rather, we report model performance averages across cross-validation folds.

Responses were tokenized using the BERT Wordpiece tokenizer with a maximum token length of 256. The tokenization process included truncation and padding to standardize input sizes. The BERT model was initialized using the standard \texttt{bert-base-uncased} pre-trained weights and fine-tuned for classification tasks. The training utilized the AdamW optimizer with a learning rate of 0.00002. Batches were processed using a batch size of 16, and the model was trained for five epochs per fold based on monitoring training loss.

Predictions from two GPT-based models (GPT-4 Turbo and GPT-4o) were included for each response and evaluated in the same manner and on the same observations as the BERT model. To generate GPT-based predictions, we engineered a prompt with 2-3 few-shot examples of how each category response should look. Based on these specifications the models were instructed to output a 0 or 1 depending on whether they believed a given learner response was adequate or not given these criteria. Unlike the BERT classifier, this approach does not allow for an output of a probability. Hence, AUC (area under the ROC-curve) could not be defined for these models. Instead, for all models we computed, the average accuracy and $F1$ score for each prediction task across folds. Finally, for each metric and model, we created an average across all prediction tasks (average across all folds) to comprehensively compare each model.

The human annotation rubrics, dataset, lesson content, generative AI model prompts, and code for the BERT model, a contribution of this work, can be found at the following GitHub repository: \href{https://github.com/conradborchers/bert-llm-open-response}{https://github.com/conradborchers/bert-llm-open-response}.

\vspace{-0.4cm}
\section{Results}

RQ1 related to the comparative performance of BERT, GPT-4o, and GPT-turbo in assessing learners' open responses within scenario-based equity training. The results presented in Table 1 demonstrate that the BERT model, fine-tuned on about 100 examples, consistently outperformed both few-shot prompted GPT models across all four prediction tasks. The Area Under the Curve (AUC) scores generally fell within a satisfactory to excellent range, spanning approximately 0.8 to 0.9. In contrast, the accuracy of the GPT models proved less consistent, with GPT-4.0 achieving higher accuracy in three out of four prediction tasks.

\begin{table}[htpb]
\caption{Comparative Performance between GPT and BERT models.}
\hspace{-0.5cm}
\begin{tabular}{|l|l|l|l|l|}
\hline
\textbf{Model} & \textbf{Task}        & \textbf{Accuracy} & \textbf{F1} & \textbf{AUC} \\ \hline
\textbf{BERT}                          & Explain - \textit{Avoiding Unconscious Assumptions} & 0.854                                     & 0.903                                     & 0.8779                               \\ \hline
\textbf{GPT-4o}                        & Explain - \textit{Avoiding Unconscious Assumptions} & 0.550                                     & 0.523                                     & n/a                                  \\ \hline
\textbf{GPT-4 Turbo}                   & Explain - \textit{Avoiding Unconscious Assumptions} & 0.798                                     & 0.805                                     & n/a                                  \\ \hline
\textbf{BERT}                          & Predict - \textit{Avoiding Unconscious Assumptions} & 0.720                                     & 0.797                                     & 0.7936                               \\ \hline
\textbf{GPT-4o}                        & Predict - \textit{Avoiding Unconscious Assumptions} & 0.754                                     & 0.755                                     & n/a                                  \\ \hline
\textbf{GPT-4 Turbo}                   & Predict - \textit{Avoiding Unconscious Assumptions} & 0.506                                     & 0.480                                     & n/a                                  \\ \hline
\textbf{BERT}                          & Explain - \textit{Helping Students Manage Inequity} & 0.896                                     & 0.909                                     & 0.9527                               \\ \hline
\textbf{GPT-4o}                        & Explain - \textit{Helping Students Manage Inequity} & 0.868                                     & 0.866                                     & n/a                                  \\ \hline
\textbf{GPT-4 Turbo}                   & Explain - \textit{Helping Students Manage Inequity} & 0.598                                     & 0.526                                     & n/a                                  \\ \hline
\textbf{BERT}                          & Predict - \textit{Helping Students Manage Inequity} & 0.880                                     & 0.935                                     & 0.8573                               \\ \hline
\textbf{GPT-4o}                        & Predict - \textit{Helping Students Manage Inequity} & 0.733                                     & 0.737                                     & n/a                                  \\ \hline
\textbf{GPT-4 Turbo}                   & Predict - \textit{Helping Students Manage Inequity} & 0.293                                     & 0.390                                     & n/a                                  \\ \hline
\textbf{BERT}                          & Average Across Tasks                         & 0.837                                     & 0.886                                     & 0.8704                               \\ \hline
\textbf{GPT-4o}                        & Average Across Tasks                         & 0.726                                     & 0.720                                     & n/a                                  \\ \hline
\textbf{GPT-4 Turbo}                   & Average Across Tasks                         & 0.549                                     & 0.550                                     & n/a                                  \\ \hline
\end{tabular}
\end{table}

$F1$ scores indicated moderate to satisfactory performance for GPT-4.0, except for the task related to explaining how to avoid unconscious assumptions. Overall, few-shot prompting with the newer and commercial GPT architectures did not surpass the performance of the resource-efficient and offline BERT fine-tuning approach on these four prediction tasks.
\vspace{-0.9cm}
\section{Discussion}

Advances in LLMs promise to open the door to assessing increasingly complex and nuanced open-response assessment tasks in education. However, our results showcase that fine-tuning early-generation models such as BERT not only led to superior accuracy than few-shot LLM prompting in the ill-defined domain of equity training, but is also more resource-efficient.
\vspace{-0.4cm}
\subsection{Efficiency and Resource Utilization}
The average performance of BERT at 0.837 was higher than both GPT models at 0.726 and 0.549 respectively. BERT demonstrated competitive performance using an offline approach that is more resource-efficient compared to deploying commercial GPT models. Moreover, BERT might be more stable in performance, too, because it does not face the continuous new model versioning often common for closed-source models such as GPT, which may alter model behavior and performance on specific tasks.
    
    The lower performance of the GPT models across both tasks could be attributed to these models primarily relying on few-shot prompt engineering techniques to assess responses. This approach does reduce the need for constant retraining; however, it may struggle to adequately capture nuanced response patterns, especially in complex, multi-faceted tasks where grading is often performed according to specific rubrics. In this task, GPT's reliance on few-shot prompting relies heavily on the ability of the model to generalize given a few examples, which can increase variability within the responses in ill-defined domains. 
\vspace{-0.3cm}

\subsection{Limitations}

The lack of probability outputs from GPT models limited metric diversity, excluding AUC as a measure for GPT models as this metric relies on probabilities, not binarized predictions. Future research may use LLM embedding-based prediction models to circumvent this issue \citep{zhang2024using}, and consider using a larger set of open-source language models compared to the LLMs used here.

Another limitation is related to our sample size, which constrained our ability to evaluate model performance on hold-out test sets beyond cross-validation. Such evaluations may pose a better estimate of the generalizability of our classifiers. Our work studies higher education student populations completing equity training, however, an extended tutor sample size in different cultural and instructional contexts (i.e., hybrid learning) and more diverse samples could allow us to make more robust conclusions for this form of the classification task.

\vspace{-0.2cm}
\subsection{Implications for Tutoring Systems}

BERT's robust performance suggests its potential for integration into assessment and learning systems for automated evaluation of learner responses at scale. Applications of such assessment models are manifold but include real-time evaluation of tutorial dialog during problem-solving practice, which may enhance learning \citep{borchers2024combining}. The ability of these models to grade open responses underscores their potential for use in challenging educational domains, such as equity training studied here.

GPT models, while promising in some tasks, may require additional adaptation or fine-tuning to improve consistency and task-specific performance compared to relying solely on few-shot examples and explanations that carry significant variability.

\vspace{-0.3cm}
\section{Future Work and Conclusion}

This study put the abilities of contemporary LLMs to the test of understanding nuanced situations of reacting to students during tutoring in an equitable fashion. Our findings suggest that the assessment task studied here requires fine-tuning beyond few-shot prompting, whereby fine-tuned BERT demonstrated superior performance across all prediction tasks across the two lessons. This performance emphasizes the value of fine-tuning task-specific datasets, even with small sample sizes. These results underscore the limitations of few-shot prompting for GPT models in handling tasks within ill-defined educational domains that require nuanced understanding and contextual reasoning, such as equity training. Resource-efficient, offline methods like BERT fine-tuning, on the other hand, offer a viable, open-source, and cost-efficient alternative to commercial AI models for educational applications.

Future work should explore increasing dataset sizes, enhancing GPT model prompt engineering approaches, exploring costs related to different model sizes and computation, and evaluating other fine-tuning strategies (such as fine-tuning GPT) to improve model performance on complex tutoring tasks. Our approach comparing GPT performance to BERT can also be extended to various other lessons on tutoring competencies, such as addressing microaggressions and narrowing opportunity gaps along with other educational domains beyond equity training. Our open availability of study materials, data, and code will facilitate replication, transparency, and future advancements in automated assessment of open-responses in scenario-based training.


\bibliography{jmlr-sample}




\end{document}